\documentclass[aps]{revtex4}
\usepackage{eurosym}
\usepackage{amsfonts}
\usepackage{amsmath}
\usepackage{amssymb,epsf}
\usepackage{color}
\usepackage{graphicx}
\usepackage{epstopdf}
\usepackage{float}
\usepackage{caption}
\usepackage{subfig}

\begin{document}

\title{Can the power Maxwell nonlinear electrodynamics theory remove the
singularity of electric field of point-like charges at their locations?}
\author{B. Eslam Panah$^{1,2,3}$\footnote{email address: eslampanah@umz.ac.ir} }
\affiliation{$^{1}$ Department of Theoretical Physics, Faculty of Basic Sciences, University of Mazandaran, P. O. Box 47416-95447, Babolsar, Iran\\
$^{2}$ ICRANet-Mazandaran, University of Mazandaran, P. O. Box 47416-95447, Babolsar, Iran\\
$^{3}$ ICRANet, Piazza della Repubblica 10, I-65122 Pescara, Italy}

\begin{abstract}
YES! We introduce a variable power Maxwell nonlinear electrodynamics theory
which can remove the singularity of electric field of point-like charges at
their locations. One of the main problems of Maxwell's electromagnetic field
theory is related to the existence of singularity for electric field of
point-like charges at their locations. In other words, the electric field of
a point-like charge diverges at the charge location which leads to an
infinite self-energy. In order to remove this singularity a few nonlinear
electrodynamics (NED) theories have been introduced. Born-Infeld (BI) NED
theory is one of the most famous of them. However the power Maxwell (PM) NED
cannot remove this singularity. In this paper, we show that the PM NED
theory can remove this singularity, when the power of PM NED is less than $s<\frac{1}{2}$.
\end{abstract}

\maketitle

\section{Introduction}

The nonlinear field theories are interesting issues in physics because most of
physical systems have a nonlinear behavior. As an example, the phenomenon of
vacuum polarization in quantum electrodynamics, arising from the
polarization of virtual electron-positron pairs and leading to nonlinear
interactions between electromagnetic fields (an example that illustrates
this is the scattering of photons by photons). The nonlinear electrodynamics
(NED) theories come from the fact that these theories are generalizations of
the linear Maxwell's theory and can remove some of problems Maxwell's theory
and also in the special case they reduce to the Maxwell's theory. Other
motivations of considering NED theories are limitations of the Maxwell
theory \cite{DelphenichI,DelphenichII}, description of the self-interaction
of virtual electron-positron pairs \cite{EPpairI,EPpairII,EPpairIII}, the
radiation propagation inside specific materials \cite%
{radiationI,radiationII,radiationIII,radiationIV}. Also NED field modifies
spacetime geometry (the gravitational red-shift) around superstrong
magnetized compact objects \cite{IbrahimI,IbrahimII,IbrahimIII}, remove both
of the big bang and black hole singularities \cite%
{singularitiesI,singularitiesII,singularitiesIII,singularitiesIV,singularitiesV}. Moreover, the effects of NED field are important for studying higher
magnetized neutron stars and pulsars \cite{Bialynicka,Mosquera}.

From a classical point of view, the existence of singularity at a point-like
charge position is one of the main problems of Maxwell's electromagnetic
field theory. In order to obtain a finite self-energy of a point charge,
Born and Infeld (BI) have introduced a NED field by modifying Maxwell's
theory at the short distance \cite{BI}. Another interesting properties of
the BI theory is that its effective action arises in an open superstring
theory and D-brains with non-singular self-energy of the point-like charge 
\cite{DBrainI,DBrainII,DBrainIII,DBrainIV}. Moreover, other NED theories
were proposed in references \cite%
{NEDI,NEDII,NEDIII,NEDIV,NEDV,NEDVI,NEDVII,NEDVIII}, where this singularity
is absent.

One of the interesting classes of the NED sources is power Maxwell (PM) NED
theory. The Lagrangian this theory is an arbitrary power of Maxwell's
Lagrangian \cite{PMI,PMII,PMIII,PMIV,PMV,PMVI}, which is invariant under the
conformal transformation $g_{\mu \nu }\rightarrow \Omega ^{2}g_{\mu \nu }$,
and $A_{\mu }\rightarrow A_{\mu }$ ($g_{\mu \nu }$ and $A_{\mu }$ are metric
tensor and electrical gauge potential, respectively). The PM NED theory is
richer than linear Maxwell's theory and in a special case (unit power),
reduces to linear Maxwell's theory \cite{PMI,PMII,PMIII,PMIV,PMV,PMVI}. In
is notable that the coupling of PM NED theory with gravity always is an
interesting subject which attracted significant attentions because of
specific properties \cite%
{Couple5,Couple6,Couple8,Couple12,Couple15,Couple16,Couple17,Couple18,Couple19,Couple20,Couple22,Couple28}. Another attractive feature of the PM NED theory is related to its conformal invariant. Indeed by adjusting the power of PM NED theory equals with a quarter of spacetime dimensions ($s=D/4$ where $D$ and $s$ are dimensions of spacetime and power of PM NED theory, respectively), the theory is conformal invariant. In other words, for the special choice of $s=D/4$, one obtains traceless energy-momentum tensor which leads to conformal invariant. It is notable that the idea is to take advantages of the conformal symmetry to construct the analogues of four-dimensional\ Reissner-Nordstr\"{o}m solution with an inverse square electric field in arbitrary dimensions ($E\propto\frac{1}{r^{2}}$, where $E$ is the electric field) \cite{PMI,CPMI,CPMII,CPMIII}. But the PM NED theory suffers from a main problem similar to linear Maxwell's theory and it is related to the existence of singularity for electromagnetic field of point-like charges at their locations. In order to remove this problem we study the behavior of PM NED theory with large value of $s$ at position of a point-like charge or near small charged black hole. In other words, our aim is related to remove singularities electromagnetic field of point-like charges at their locations of the PM NED theory by considering a variable PM NED theory.

\section{The electric field in PM NED Theory}

The $4$-dimensional action of a NED is given by 
\begin{equation}
\mathcal{S}=\int d^{4}x\sqrt{-g}\mathcal{L}_{NED}\left( \mathcal{F}\right) ,
\label{action}
\end{equation}%
where $\mathcal{L}_{NED}\left( \mathcal{F}\right) $ is an arbitrary
Lagrangian of NED and $g:=\det \left( g_{\mu \nu }\right) $. Varying the action (\ref{action}) with respect to the gauge potential $A_{\mu }$, we obtain the field equation as 
\begin{equation}
\partial _{\mu }\left( \sqrt{-g}\mathcal{L}_{F}F^{\mu \nu }\right) =0,
\label{NEDeq}
\end{equation}%
where $\mathcal{L}_{F}=d\mathcal{L}_{NED}\left( \mathcal{F}\right) /d%
\mathcal{F}$, and $\mathcal{F}=F_{\mu \nu }F^{\mu \nu }$ is the Maxwell
invariant. Also $F_{\mu \nu }=\partial _{\mu }A_{\nu }-\partial _{\nu
}A_{\mu }$ is the electromagnetic tensor field.

It is well-known that the electric field is associated with the time
component of the vector potential ($A_{t}$). So we assume the vector
potential as $A_{\mu }=h(r)\delta _{\mu }^{0}$.

The Lagrangian of PM NED theory is $\mathcal{L}_{NED}\left( \mathcal{F}%
\right) =\left( -\mathcal{F}\right) ^{s}$. By replacing the Lagrangian  of PM NED theory in Eq. (\ref{NEDeq}), we have
\begin{equation}
\partial _{\mu }\left( \sqrt{-g}\left( -\mathcal{F}\right) ^{s-1}F^{\mu \nu
}\right) =0.  \label{PMeq}
\end{equation}

Using the electrical gauge potential ($A_{\mu }=h(r)\delta _{\mu }^{0}$) and
Eq. (\ref{PMeq}), and by considering a $4$-dimensional static spherical
symmetric spacetime as $ds^{2}=-dt^{2}+dr^{2}+r^{2}\left(
d\theta ^{2}+\sin ^{2}\theta d\varphi ^{2}\right) $, we obtain 
\begin{equation}
rh\left( r\right) {^{\prime \prime }}\left( 2s-1\right) +2h\left( r\right) {%
^{\prime }}=0,
\end{equation}%
where the prime and the double prime denote the first and second derivatives
with respect to $r$. Using the above equation we can extract the electrical
gauge potential in the following form 
\begin{equation}
h\left( r\right) =qr^{\frac{2s-3}{2s-1}},  \label{hr}
\end{equation}%
where $q$ is an integration constant which is related to the electric charge.

Now we investigate the physical limit of power of the PM NED theory by
evaluating the electrical gauge potential both at far from of the source ($%
r\rightarrow \infty $) and at origin or at very short distance of the source
($r\rightarrow 0$).

For $r\rightarrow \infty $: the electrical gauge potential should be finite.
Therefore, one should impose a restriction on $s$ as $\frac{2s-3}{2s-1}<0$,
which leads to a restriction in the following form $\frac{1}{2}<s<\frac{3}{2}
$.

For $r\rightarrow 0$: by considering $s<\frac{1}{2}$, we can remove the
singularity of electrical gauge potential of a point charge at its location.
In other words, for $s<\frac{1}{2}$, $h(r)=qr^{\varepsilon }$ ($\varepsilon $
is positive value), which indicate that the electrical gauge potential is
finite when $r\rightarrow 0$.

So the physical limit of $s$ for the electrical gauge potential is as 
\begin{equation}
\text{Physical limit for }h\left( r\right) \text{ leads to }\rightarrow
\left\{ 
\begin{array}{ccc}
\frac{1}{2}<s<\frac{3}{2} & \text{for} & r\rightarrow \infty  \\ 
&  &  \\ 
s<\frac{1}{2} & \text{for} & r\rightarrow 0%
\end{array}%
\right. .
\end{equation}

Using Eq. (\ref{hr}), one can extract the electric field of PM NED theory as 
\begin{equation}
E\left( r\right) =\frac{dh(r)}{dr}=qr^{\frac{-2}{2s-1}}.  \label{Er}
\end{equation}

\begin{figure*}[tbh]
\centering
\includegraphics[width=0.45\linewidth]{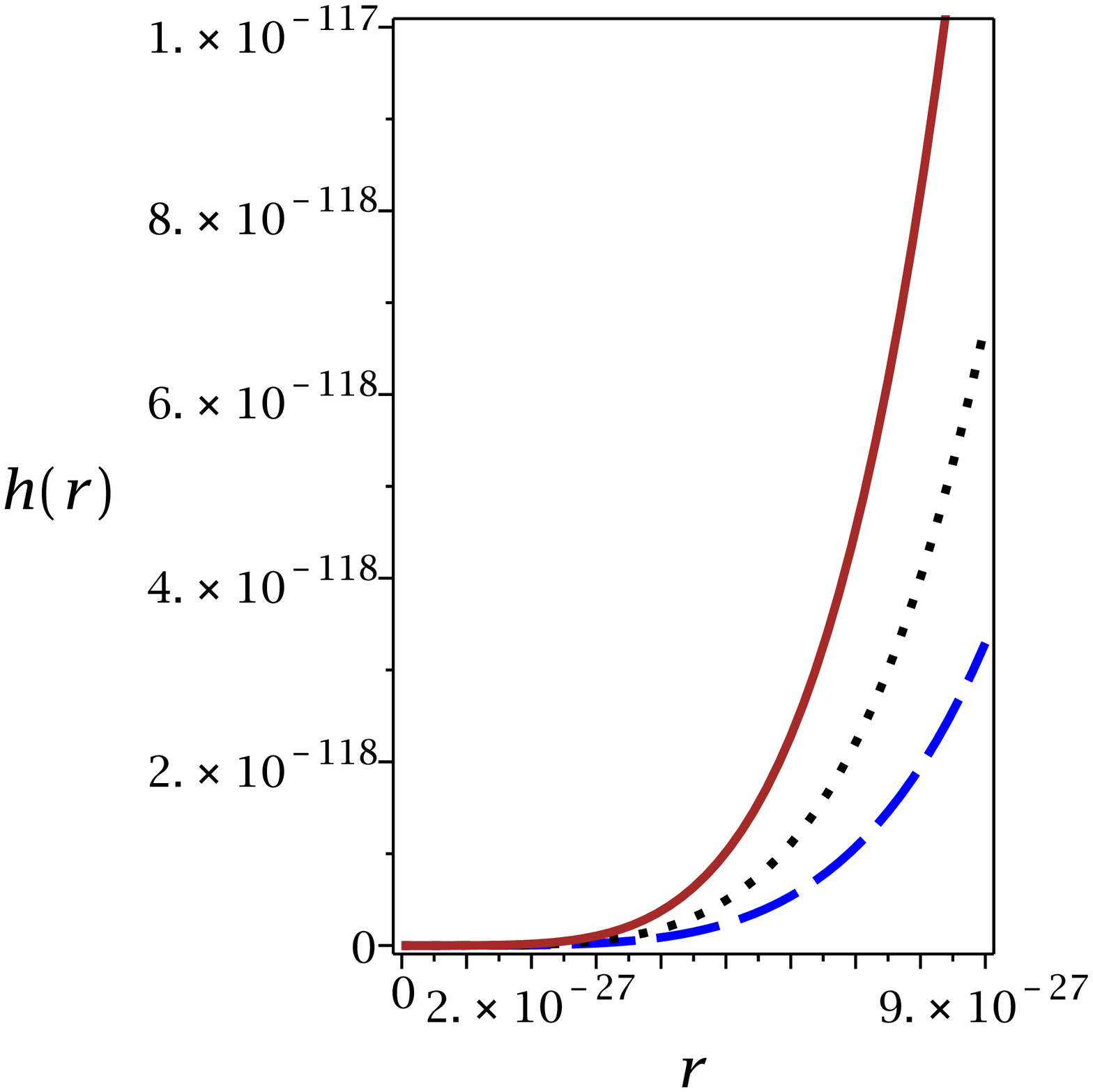} \includegraphics[width=0.45%
\linewidth]{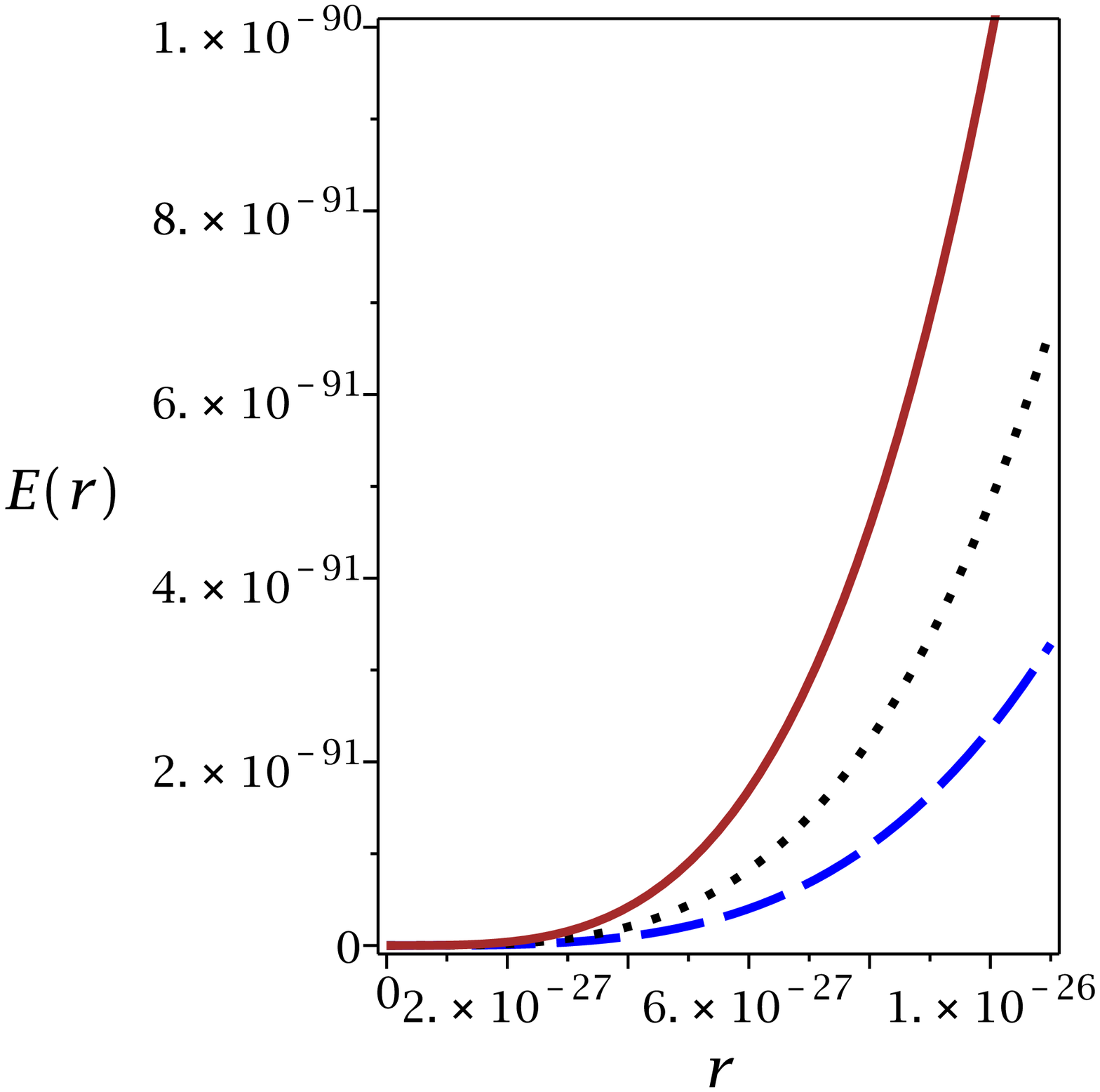}
\caption{$h(r)$ (left panel) and $E(r)$ (right panel) of point-like
	charges at near their locations versus $r$, for $q=0.1$, $s=0.212$ (dashed
line), $s=0.211$ (dotted line) and $s=0.210$ (continuous line).}
\label{Fig1}
\end{figure*}

\begin{figure*}[tbh]
	\centering
	\includegraphics[width=0.5\linewidth]{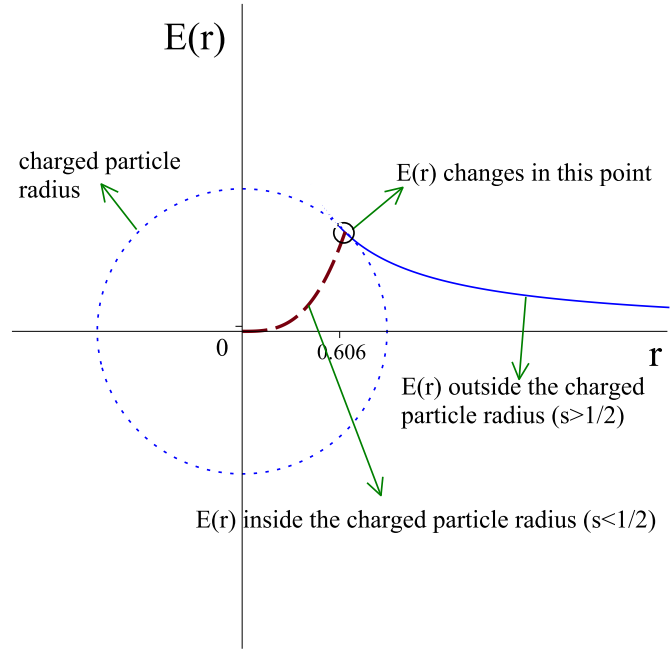} 
	\caption{$E(r)$ versus $r$, inside ($s<1/2$) and outside ($s>1/2$) of the charged particle radius.}
	\label{Fig2}
\end{figure*}

We can remove the singularity of electrical field of a point charge at its
location, when that $s$ is less than $\frac{1}{2}$ $\left( s<\frac{1}{2}%
\right) $. In this case, the electrical field of PM NED theory turns to 
\begin{equation}
E\left( r\right) =qr^{\varepsilon },
\end{equation}%
where indicates the electrical field of point-charge in PM NED theory is a decreasing function of $r$ (for more details see Figs. \ref{Fig1} and \ref{Fig2}).

On the other hand, the electrical charge should be zero at infinity distance
of source ($r\rightarrow \infty $). This fact impose a restriction on the
range of $s$ as $s>\frac{1}{2}$. Therefore, we found the physical limit of $%
s $ for the electrical field as

\begin{equation}
\text{Physical limit for }E\left( r\right) \text{ leads to }\rightarrow
\left\{ 
\begin{array}{ccc}
s>\frac{1}{2} & \text{for} & r\rightarrow \infty  \\ 
&  &  \\ 
s<\frac{1}{2} & \text{for} & r\rightarrow 0%
\end{array}%
\right. .
\end{equation}

See Table. \ref{tab1}, for more details about acceptable ramges of $s$ based
on the physical limit of the electrical gauge potential and the electric
field.

So the PM NED theory is a theory without any singularities of the electric
field of point-like charges at their locations when $s$ is less than $\frac{1%
}{2}$ $\left( s<\frac{1}{2}\right) $.

In this paper we indicated that the PM NED theory can avoid of singularities
of the electrical gauge potential and the electric field of point-like
charges at their locations, when $s<\frac{1}{2}$. 
\begin{table*}[tbp]
\caption{Acceptable ranges of $s$ based on the physical limit of the
electrical gauge potential and the electric field.}
\label{tab1}
\begin{center}
\begin{tabular}{||c|c|c||c|c||}
\hline\hline
different fields & $h(r)$ when $r\rightarrow \infty$ & $h(r)$ when $r\rightarrow 0$ & $E(r)$
when $r\rightarrow \infty$ & $E(r)$ when $r\rightarrow 0$ \\ \hline\hline
acceptable ranges of $s$ & $\frac{1}{2}<s<\frac{3}{2}$ & $s<\frac{1}{2}$ & $s>\frac{1}{2}$ & $s<\frac{1}{2}$ \\ \hline\hline
\end{tabular}%
\end{center}
\end{table*}

\section{Conclusion}

In this paper, we have introduced a variable PM NED theory which can remove
the singularity of electric field of point-like charges at their locations,
when $s<\frac{1}{2}$. Indeed, we have indicated that the electrical field of
a point-like charge in the variable PM NED theory is finite at $r\rightarrow 0$. Also, we have obtained acceptable ranges of power of PM NED ($s$) for all spacetime in Table. \ref{tab1}.


\end{document}